\documentclass[3p,times]{elsarticle}

\usepackage{color}
\usepackage{amsmath}  
\usepackage{amsfonts} 
\usepackage{graphicx} 
\usepackage{hyperref}

\usepackage{times}
\usepackage{bm}
\usepackage{slashed}

\newcommand{\pv}{\mathbf{p}}
\newcommand{\xv}{\mathbf{x}}
\newcommand{\er}{\mathrm{e}}

\newcommand{\rv}{\mathbf{r}}
\newcommand{\wv}{\mathbf{w}}
\newcommand{\cv}{\mathbf{c}}
\newcommand{\uv}{\mathbf{u}}
\newcommand{\Imc}{I_{\mathrm{MC}}}
\newcommand{\sigmc}{\sigma_{\mathrm{MC}}}

\newcommand{\red}[1]{{\color{red} #1}}
\renewcommand{\red}[1]{{ #1}}  

\newcommand{\Eq}[1]{Eq.~(\ref{eq:#1})}

\newcommand{\vegas}{\textsc{vegas}}
\newcommand{\vegasp}{\textsc{vegas+}}
\newcommand{\miserp}{\textsc{miser+}}
\newcommand{\miser}{\textsc{miser}}
\newcommand{\tq}{\textsc{tq}s}
\newcommand{\Nev}{N_\mathrm{ev}}
\newcommand{\Ng}{N_g}
\newcommand{\Nit}{N_\mathrm{it}}
\newcommand{\Nst}{N_\mathrm{st}}
\newcommand{\onlinecite}[1]{Ref.~\cite{#1}}

\begin{document}

\title{Adaptive  Multidimensional Integration: \vegas\ Enhanced}

\author{G. Peter Lepage}
\ead{g.p.lepage@cornell.edu} 
\address{Department of Physics, Cornell University, Ithaca, NY, 14853}

\date{\today}

\begin{abstract}
We describe a new algorithm, \vegasp, for adaptive multidimensional Monte Carlo integration. The new algorithm adds a second adaptive strategy, adaptive stratified sampling, to the adaptive importance sampling that is the basis for its widely used predecessor~\vegas. Both \vegas\ and \vegasp\ are effective for integrands with large peaks, but \vegasp\ can be much more effective for integrands with multiple peaks or other significant structures aligned with diagonals of the integration volume. We give examples where \vegasp\ is 2--19$\times$ more accurate than \vegas. We also show how to combine \vegasp\ with other integrators, such as the widely available \miser\ algorithm, to make new hybrid integrators. For a different kind of hybrid, we show how to use integrand samples, generated using MCMC or other methods, to optimize \vegasp\ before integrating. We give an example where preconditioned \vegasp\ is more than 100$\times$ as efficient as \vegasp\ without preconditioning. Finally, we give examples where \vegasp\ is more than 10$\times$ as efficient as MCMC for Bayesian integrals with~$D=3$ and~21 parameters. We explain why \vegasp\ will often outperform MCMC for small and moderate sized problems.
\end{abstract}
\maketitle

\section{Introduction}
Classic \vegas\ is an algorithm for  adaptive  multidimensional Monte Carlo integration~\cite{vegas}. It is widely used in particle physics: for example, in Monte Carlo event generators~\cite{event}, and to evaluate low-order~\cite{low-order} and high-order~\cite{high-order} Feynman diagrams and cross sections numerically. It has also been used in other fields for a variety of applications including, for example, path integrals for chemical physics~\cite{quantum-gases} and option pricing in applied finance~\cite{options}, Bayesian statistics for astrophysics~\cite{supernova,tidal,bulges} and medical statistics~\cite{bioterror}, models of neuronal networks~\cite{neuronal},  wave-function overlaps for atomic physics~\cite{overlap}, topological integrals for condensed matter physics~\cite{topological}, and so on. 

Monte Carlo integration is unusually robust. It makes few assumptions about the integrand\,---\,it needn't be analytic or even continuous\,---\,and provides useful measures for the uncertainty and reliability of its results.  This makes it well suited to multidimensional integration, and in particular adaptive multidimensional integration. Adaptive strategies are essential for integration in high dimensions because important structures in integrands often occupy small fractions of the integration volume. One might not think, for example, of the interior of a sphere of radius~0.5 enclosed by a unit hypercube as a ``sharp peak,'' but it occupies only 0.0000025\% of the hypercube in $D=20$~dimensions.\footnote{\vegasp, the algorithm described in this paper, can  find the 20-D sphere with 10 iterations of $\Nev=10^7$~integrand samples each. Another 10~iterations gives an estimate for the volume that is accurate to~0.05\%. }

Classic \vegas\ is very effective for integrands with sharp peaks, which are common in particle physics applications and Bayesian integrals. While it is particularly effective for integrals that are separable (into a product of one-dimensional integrals), it also works very well for non-separable integrals with large peaks. It works less well for integrands with multiple peaks or other important structures aligned with diagonals of the integration volume, although it is   much better than Simple Monte Carlo integration.

In this paper, we describe a modification of classic \vegas, which we call \vegasp, that adds a second adaptive strategy to classic \vegas's adaptive \emph{importance sampling}~\cite{oldref}. The second strategy is a form of adaptive \emph{stratified sampling}~\cite{oldref} that makes \vegasp\ far more effective in dealing with multiple peaks and diagonal structures in integrands, given enough integrand samples. 

We begin in Sec.~\ref{sec:adaptive_importance_sampling} by reviewing the  algorithm for (iterative) adaptive importance sampling used in classic \vegas. We discuss in particular the variable map used by \vegas\ to switch to new integration variables that flatten the integrand's peaks. This is how \vegas\ implements importance sampling. We describe the new algorithm \vegasp\ in Sec.~\ref{sec:adaptive_stratified_sampling} and give examples that illustrate how and why it works. We also discuss its limitations and how they can be mitigated.

The map used by \vegas\ to implement importance sampling can be used in conjunction with other integration algorithms to create new hybrid algorithms. We show how this is done in Sec.~\ref{sec:vegasp_hybrids}, where we combine \vegasp\ with the \miser\ algorithm~\cite{miser}. There we also show how to use sample data, from a Markov Chain Monte Carlo (MCMC) or other peak-finding algorithm, to optimize the \vegas~map before integrating; this can greatly reduce the cost of integrating functions with multiple high, narrow peaks.

Our conclusions are in Sec.~\ref{sec:conclusions}. We continue in two appendices with further \vegasp\ examples that illustrate its use for adaptive multidimensional summation, high-order Feynman diagrams, and Bayesian curve fitting, where we compare \vegasp\ with MCMC for a 3-dimensional and a 21-dimensional problem.

When comparing algorithms we do not look at computer run times because these are too sensitive to implementation details. For realistic applications the cost of evaluating the integrand usually exceeds other costs, so we measure efficiency by the number of integrand evaluations used. Where statistical uncertainties are quoted, they correspond to one standard deviation;\footnote{Error estimates from \vegas\ and \vegasp\ in most examples come from multiple iterations which were combined as described in Sec.~\ref{sec:combining}, with good $\chi^2$s in each case. Estimated values for the integrals agreed with the exact values to within errors (i.e., mostly within $\pm1\sigma$).} small differences (e.g., 10\%) between uncertainties are not significant. The \vegasp\ examples here were analyzed using a widely available Python/Cython implementation first released in~2013~\cite{vegas-code}. The integrands are written in (vectorized) Python or, in one case, Fortran77. 

One change since classic \vegas\ was introduced 45~years ago is the enormous increase in speed of computers. As a result \vegasp\ integrals using~$10^8$ or~$10^9$ integrand evaluations require only minutes for simple integrands on a modern laptop. \vegasp\ is also easily configured for parallel computing~\cite{parallel}.

\section{Adaptive Importance Sampling}
\label{sec:adaptive_importance_sampling}
In this section we review the adaptive strategy used in the original \vegas\ algorithm: 
its remapping of the integration variables in each direction to optimize Monte Carlo estimates of the integral. The strategy is an implementation of the standard technique of {importance sampling.}
In the following sections we review how \vegas\ implements this strategy, first for one-dimensional integrals and then for multidimensional integrals.

\subsection{Remapping the Integration Variable}
For one-dimensional integrals, the \vegas\ strategy is to 
replace the original integral 
\begin{equation}
    I = \int\displaylimits_a^b \! dx\,f(x) 
\end{equation}
by an equivalent integral over a new variable~$y$,
\begin{equation}
    I = \int\displaylimits_0^1\!dy\,J(y) f(x(y)),
    \label{eq:Iy}
\end{equation}
where $J(y)$ is the Jacobian of the transformation. The transformation is chosen to minimize the uncertainty in a Monte Carlo evaluation of the integral in $y$-space.

A Simple Monte Carlo estimate of the $y$-integral (\Eq{Iy}) is given by
\begin{equation}
    I \approx \Imc \equiv \frac{1}{\Nev}\sum_y J(y) f(x(y))
    \label{eq:MCIy}
\end{equation}
where the sum is over $\Nev$ random points~$y$ uniformly distributed between~0 and~1. The estimate~$\Imc$ is itself a random number whose mean is the exact value~$I$ of the integral and whose variance is 
\begin{align}
    \sigma_I^2 &= \frac{1}{\Nev}\Big(
        \int\displaylimits_0^1\!dy\,J^2(y) f^2(x(y)) - I^2 
    \Big) \\
    &= \frac{1}{\Nev}\Big(\int\displaylimits_a^b\!dx\, J(y(x)) f^2(x) - I^2
    \Big).
    \label{eq:MCsigIy}
\end{align}
The standard deviation~$\sigma_I$ is an indication of the possible error in the Monte Carlo estimate. The variance~$\sigma_I^2$ can also be estimated from the Monte Carlo data:
\begin{equation}
    \sigmc^2 \equiv \frac{1}{\Nev-1} \Big(
        \frac{1}{\Nev} \sum_y J^2(y) f^2(x(y)) - \Imc^2
    \Big)
    \label{eq:sigest}
\end{equation}

The distribution of the Monte Carlo estimates~$\Imc$ becomes Gaussian in the limit of large~$\Nev$, assuming $J(y)f(x(y))$ is sufficiently integrable. There are non-Gaussian corrections, but they vanish quickly with increasing~$\Nev$. For example, the Monte Carlo estimates for~$I$ and~$\sigma_I^2$ would be uncorrelated with Gaussian statistics, but here have a nonzero correlation that vanishes like~$1/\Nev^2$:
\begin{align}
    \big\langle (\Imc - I)&(\sigmc^2 - \sigma_I^2) \big\rangle 
    = \big\langle (\Imc - I)^3 \big\rangle \nonumber \\ 
    &= \frac{1}{\Nev^2}\int\displaylimits^1_0\!dy\,\big(J(y) f(x(y)) - I\big)^3.
    \label{eq:correlation}
\end{align}

\subsection{\vegas~Map and Importance Sampling}
\vegas\ implements the variable map described in the previous section by dividing the $x$-axis into $\Ng$~intervals bounded by:
\begin{align}
    x_0 &= a  \nonumber \\
    x_1 &= x_0 + \Delta x_0 \nonumber \\ 
    \cdots \nonumber \\
    x_{\Ng} &= x_{\Ng - 1} + \Delta x_{\Ng - 1} = b.
\end{align}
where $\Ng = 1000$ is typical.
The transformed variable has value $y=i/\Ng$ at point~$x_i$, and varies linearly with~$x$ between~$x_i$s. Thus 
\begin{equation}
    x(y) \equiv x_{i(y)} + \Delta x_{i(y)} \,\delta(y)
    \label{eq:y}
\end{equation}
relates~$x$ to~$y$, where functions $i(y)$ and $\delta(y)$ are the integer and fractional parts of~$y\Ng$, respectively:
\begin{align}
    i(y) &\equiv \mathrm{floor}(y\Ng) \\ 
    \delta(y) &\equiv y\Ng - i(y).
\end{align}
This transformation maps the interval~$[0,1]$ in $y$-space onto the the original integration region~$[a,b]$ in $x$-space. Intervals of varying widths~$\Delta x_i$ in $x$-space map into intervals of uniform width $\Delta y = 1/\Ng$ in $y$-space. The Jacobian for this transformation, 
\begin{equation}
    J(y) = \Ng \Delta x_{i(y)} \equiv J_{i(y)},
    \label{eq:jac}
\end{equation}
is a step function whose values are determined by the interval widths~$\Delta x_i$.

Substituting this Jacobian into \Eq{MCsigIy} for the uncertainty in a Monte Carlo integration gives 
\begin{equation}
    \sigma_I^2 = \frac{1}{\Nev}\Big(
        \sum_i J_i \! \int\displaylimits_{x_i}^{x_{i} + \Delta x_i} \!\! dx\, f^2(x) - I^2
        \Big).
\end{equation}
Treating the $J_i$ as independent variables, subject to the constraint
\begin{equation}
    \sum_i \frac{\Delta x_i}{J_i} = \sum_i \Delta y = 1,
\end{equation}
it is easy to show that $\sigma_I^2$ is minimized when 
\begin{align}
    \frac{J_i^2}{\Delta x_i} \int\displaylimits^{x_{i} + \Delta x_i}_{x_i}\!\!dx\, f^2(x)
    = \mbox{constant}.
    \label{eq:onedopt}
\end{align}
That is, the grid is optimal when the average value of $J^2(y(x)) f^2(x)$ in each interval~$\Delta x_i$ is the same for every interval. 

A transformation with this property greatly reduces the standard deviation when the integrand has high peaks. The Jacobian flattens the peaks in $y$-space and stretches them out, because
\begin{equation}
    J \equiv \left|\frac{dx}{dy}\right| \propto \frac{1}{|f(x)|}
    \label{eq:invJ}
\end{equation}
becomes small near a peak. 
This means that a uniform Monte Carlo in $y$-space concentrates integrand samples around the peaks in $x$-space. Each interval~$\Delta x_i$ receives on average the same number of samples ($=\Nev/\Ng$), and the smallest intervals are placed where $|f(x)|$~is largest (because $J_i\propto\Delta x_i$). This concentrates samples in the most important regions, which is why this method is called {importance sampling}.

\subsection{Iterative Adaptation}
\label{sec:iterative-adaptation}
The set of $x_i$s defined above constitutes the \vegas~map. To optimize Monte Carlo integration, \vegas\ varies the Jacobian of the transformation by varying the interval sizes~$\Delta x_i$, while keeping the sum of all~$\Delta x_i$s constant. This is done iteratively. First \vegas\ estimates the integral with a uniform grid, accumulating information about the integrand in the process. This information is then used to construct an improved grid, and \vegas\ makes a new estimate of the integral. Again information about the integrand is accumulated in the process, and used to further improve the grid. In this fashion, the \vegas~map adapts to the integrand over several iterations. 

To illustrate how this is done, we continue with the example above where we generate a Monte Carlo estimate of the integral in $y$-space (\Eq{Iy}) by sampling the integrand at $\Nev$~random  points~$y$ (\Eq{MCIy}). Given some initial grid, \vegas\ accumulates the average value of $J^2f^2$ for each interval in the grid while sampling the integrand to estimate the integral:
\begin{equation}
    d_i \equiv 
    \frac{1}{n_i}   
    \sum_{x(y)\in \Delta x_i} \! J^2(y) f^2(x(y)),
\end{equation} 
where $n_i\approx\Nev/\Ng$ is the number of samples in interval~$\Delta x_i$. The averages $d_i$ are used to refine the grid. The grid is optimal when all of the $d_i$s are equal (\Eq{onedopt}), so \vegas\ adjusts the grid intervals~$\Delta x_i$ to make the $d_i$ more constant across the integration region. 

This algorithm, like most adaptive algorithms, tends to overreact in the early stages of optimizing its grid, since it has rather poor information concerning the integrand  at this stage. It is important, therefore, to dampen the refinement process so as to avoid rapid, destabilizing changes in the grid. In \vegas, the $d_i$s are first smoothed and normalized:
\begin{equation}
    d_i \to \frac{1}{\sum_i d_i}\times 
        \begin{cases}
        (7 d_0 + d_1) / 8  & \mbox{for $i=0$} \\
        (d_{i-1} + 6 d_i + d_{i+1}) / 8  & \mbox{for $i\ne 0,\Ng-1$} \\
        (d_{\Ng-2} + 7 d_{\Ng-1})/8 & \mbox{for $i=\Ng-1$}.
        \end{cases}
\end{equation}
Smoothing is particularly important if the integrand has large discontinuities (e.g., step functions). For example, an abrupt increase in the integrand near the upper edge of interval~$\Delta x_i$ might be missed completely by the samples. Without smoothing, \vegas\ would see a sudden rise in the function beginning only in~$\Delta x_{i+1}$, and refine the grid accordingly, thereby missing the small but possibly significant part of the step in interval~$\Delta x_i$. Smoothing makes this less likely by causing \vegas\ to focus some effort on interval~$\Delta x_i$ as well as~$\Delta x_{i+1}$.

Having smoothed the  $d_i$s, \vegas\ then compresses their range, to avoid overreacting to atypically large sample values for the integrand. This is done by replacing each $d_i$ with
\begin{equation}
    d_i \to \Big(\frac{1 - d_i}{\mathrm{ln}(1/d_i)}\Big)^\alpha ,
\end{equation}
where $\alpha\ge0$ is typically of order one.\footnote{Classic \vegas\ typically defaults to $\alpha=1.5$. \vegasp\ can use a smaller default value, $\alpha=0.5$, because of its adaptive stratified sampling.} Parameter~$\alpha$ can be reduced in situations where \vegas\ has trouble finding or holding onto the optimal grid; $\alpha=0$ implies no grid refinement.

The condition for an optimal map remains that all $d_i$, now smoothed and compressed, be roughly equal. If the map is not optimal, \vegas\ attempts to improve it. First the $d_i$s are treated as continuous quantities, and each~$d_i$ is distributed uniformly over its interval~$\Delta x_i$. Then new intervals, specified by $\{x_i^\prime, \Delta x_i^\prime\}$, are chosen so that each contains an equal fraction of the total~$d =\sum_i d_i$. The following algorithm achieves this:
\begin{enumerate}
    \item Define $\delta d$ to be the amount of $d$ associated with each interval of the new grid,
    \begin{equation}
        \delta d \equiv \frac{\sum_i d_i}{\Ng},
    \end{equation}
    and initialize the following variables:
    \begin{align}
        x_0^\prime &= x_0 \nonumber \\
        x_{\Ng}^\prime &= x_{\Ng} \nonumber \\
        i &= 0 = \mbox{index of current new $x^\prime$} \nonumber \\
        j &= 0 = \mbox{index of current old $x$} \nonumber \\
        S_d &= 0 = \mbox{amount of $d$ accumulated}.
    \end{align}

    \item Increment~$i$. If $i\ge\Ng$, the new grid is finished.

    \item Skip to the next step if $S_d\ge \delta d$; otherwise add $d_j$ to $S_d$, increment~$j$, and return to the beginning of this step.
    
    \item Subtract $\delta d$ from $S_d$, and compute the boundary of the new interval by interpolation:
    \begin{equation}
        x_i^\prime = x_j - \frac{S_d}{d_{j-1}} \Delta x_{j-1}.
    \end{equation}
    Return to Step~2.
\end{enumerate}
Replacing the old map with the new map, \vegas\ proceeds to generate a new 
Monte Carlo estimate for the integral and another new map. This entire 
process is repeated until the map has converged and the estimate of the integral is sufficiently accurate.

\subsection{Multidimensional Integrals}
\label{sec:multidimensional}
Monte Carlo integration, even with adaptive importance sampling, is not usually competitive with other algorithms for one-dimensional integration. It rapidly becomes competitive, however, as the dimensionality increases. 

The \vegas\ algorithm extends the algorithm described above to $D$-dimensional integrals over variables~$x^\mu$ (with $\mu=1\ldots D$) by replacing each~$x^\mu$ with a new variable~$y^\mu$. The variable transformation is specified by an independent \vegas~map $\{x_i^\mu,\Delta x_i^\mu\}$ for each direction. 

The resulting integral is 
\begin{equation}
    I = \int\displaylimits_0^1\!d^Dy\,J(y)\,f(x(y))
    \label{eq:y-Dspace}
\end{equation}
where $x(y)$ is the $D$-dimensional \vegas~map and $J(y)$ its Jacobian.
A Simple Monte Carlo estimate of this integral is obtained by sampling the integrand at random points~$y=\{y^\mu\}$ distributed uniformly within the unit hypercube at the origin:
\begin{equation}
    0 < y^\mu < 1.
\end{equation}
The integrand samples are also used to calculate the averages
\begin{equation}
    d_i^\mu \equiv \frac{1}{n_i^\mu} 
    \sum_{x^\mu(y^\mu)\in \Delta x_i^\mu} \!
    J^2(y) f^2(x)
\end{equation}
for every interval on every integration axis. These are used to improve the grid for each variable after each iteration, following the procedure described in Section~\ref{sec:iterative-adaptation}.

\begin{figure}\begin{center}
        \includegraphics[scale=0.9]{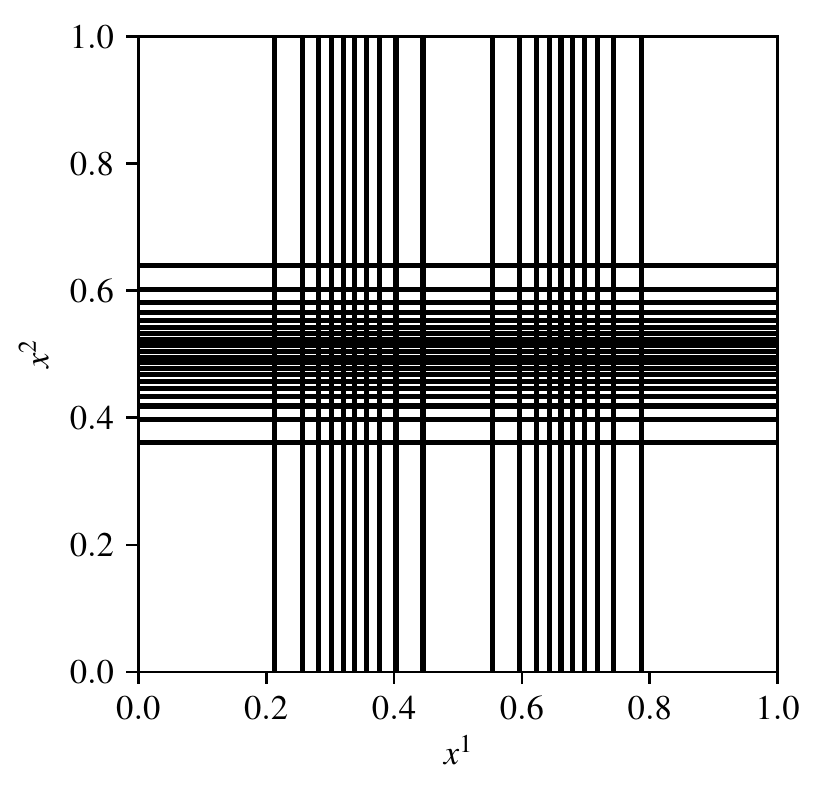}
    \caption{\label{fig:grid} \vegas~map for the $D=4$ integral defined in Eqs.~(\ref{eq:easy}) and~\Eq{easyr}. The figure shows grid lines for every 50\textsuperscript{th}~increment along the $x^1$ and $x^2$~axes; grids for the $x^3$ and~$x^4$ axes are the same as for~$x^2$.}
\end{center}\end{figure}

\begin{figure}\begin{center}
    \includegraphics[scale=0.9]{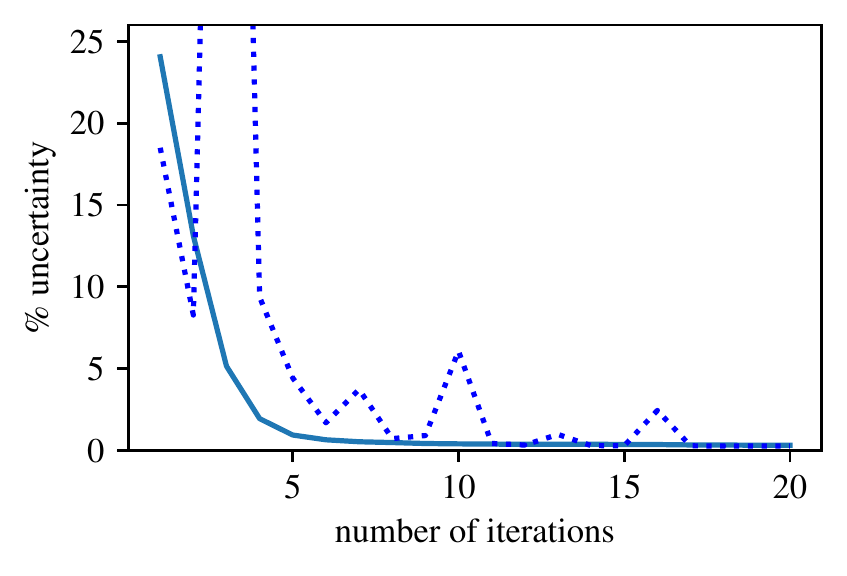}
    \caption{\label{fig:balls} Percent uncertainty ($1\sigma$) for each \vegas\ iteration in a simple $y$-space Monte Carlo for the $D=4$~integral defined in Eqs.~(\ref{eq:balls}) and~(\ref{eq:easyr}). The solid line shows how the uncertainty improves with successive iterations of the refinement process when damping parameter~$\alpha=0.2$; the dotted line shows (unstable) improvement when~$\alpha=1.0$.}
\end{center}\end{figure}

Fig.~\ref{fig:grid} shows the grid corresponding to a \vegas~map optimized for the $D=4$~dimensional integral
\begin{equation}
    \int\displaylimits^1_0 \! d^4x\,
    \Big(
        \er^{-100 (\xv - \rv_1)^2} + \er^{-100 (\xv - \rv_2)^2} 
    \Big),
    \label{eq:easy}
\end{equation}
where vector $\xv=(x^1, x^2, x^3, x^4)$, and 
\begin{align}
    \rv_1 &= (0.33, 0.5, 0.5, 0.5) \nonumber \\ 
    \rv_2 &= (0.67, 0.5, 0.5, 0.5).
    \label{eq:easyr}
\end{align}
The grid concentrates increments near~0.33 and~0.67 for~$x^1$, and near~0.5 in the other directions. Each rectangle in the figure receives, on average, the same number of Monte Carlo integration samples. 

This \vegas~map has $\Ng=1000$ increments along each axis. The accuracy of the $y$-space integrals is typically insensitive to $\Ng$ so long as it is large enough. With $\Nev=10^4$ integrand evaluations, the accuracy improves from~11\% with $\Ng=1$ (i.e., no \vegas~map) to 0.3\% at $\Ng=100$ and flattens out at~0.1\% around $\Ng=700$.

The \vegas~map is particularly effective for integrals like that in \Eq{easy}, because the integral over each Gaussian can be separated into a product of one-dimensional integrals over each direction. It also works well, however, for other integrands with high peaks that are not separable in this way.
The solid line in Fig.~\ref{fig:balls}, for example, shows how the uncertainty in a $y$-space Monte Carlo is reduced over 20~iterations for the~$D=4$ integral 
\begin{equation}
    \int\displaylimits^1_0 \! d^4x\,
    \sum_{i=1}^2
        \Theta(|\xv - \rv_i| < 0.067),
    \label{eq:balls}    
\end{equation}
where the $\rv_i$ are given in \Eq{easyr} and
\begin{equation}
    \Theta(x) = \begin{cases}
        1 & \mbox{if $x=\mathrm{True}$} \\
        0 & \mbox{if $x=\mathrm{False}$}.
    \end{cases}
\end{equation}
This integral is harder than the previous one, because the peaks have no shoulders and so are difficult to find\,---\,the integrand vanishes over 99.96\% of the integration volume. The \vegas~map is nevertheless able to reduce the fractional uncertainty from~24\% before adapting to 0.34\% after 10--20~iterations, where $\Nev=10^5$ Monte Carlo samples are used for each iteration. The damping parameter was set lower here, to~$\alpha=0.2$, to ensure smooth adaptation (solid line); setting~$\alpha=1.0$ leads to instability (dotted line). 

This last example also illustrates how the \vegas~map can improve integrals over volumes with irregular shapes. Monte Carlo integration works well even with discontinuous integrands and so has no problem with the $\Theta$ functions that define the integration volume. The \vegas~map helps find the integration region and concentrate samples in that region. Obviously it is better, where possible, to redefine the integration variables so that the integration region is rectangular. 

\subsection{Combining and Comparing Iterations}
\label{sec:combining}
\vegas, being iterative, generates a series of estimates $I_j$ of the integral (\Eq{MCIy}), each with its own error estimate $\sigma_j\equiv\sigma_{I_j}$ (\Eq{sigest}). As discussed above, the distribution of $I_j$s for a given \vegas~map is approximately Gaussian when a sufficient number of samples~$\Nev$ is used (and assuming the integral is square integrable). Then we can combine estimates from separate iterations to obtain a cumulative estimate of the integral and its standard error:
\begin{align}
    \overline I &= \frac{\sum_j I_j/\sigma^2_j}{\sum_j 1 /\sigma_j^2} \\
    \sigma_{\overline I} &= \Big(\sum_j \frac{1}{\sigma_j^2}\Big)^{-1/2}.
\end{align}
The cumulative estimate is usually superior to the estimates from separate iterations. 

It is important to verify that results from different iterations are consistent with each other within the estimated uncertainties. This provides a direct check on the reliability of the error estimates. For example, we can calculate the $\chi^2$~statistic for the estimates:
\begin{equation}
    \chi^2 = \sum_j \frac{\big(I_j - \overline I\big)^2}{\sigma_j^2}.
\end{equation}
We expect $\chi^2$ to be of order the number of iterations (less one) when the $I_j$ are approximately Gaussian, and the estimates of the uncertainties $\sigma_j$ are reliable. If $\chi^2$ is considerably larger than this, either or both of~$\overline I$ and~$\sigma_{\overline I}$ may be unreliable. There are two common causes for $\chi^2$s that are too large.

The first common cause is that early iterations, before the \vegas~map has adapted, may give very poor estimates for the integral and its uncertainty. This happens, for example, when the integrand has high, narrow peaks that are largely missed in early iterations, leading to estimates for the integral and error that are both much too small. A standard remedy is to omit the early iterations from the determinations of $\overline{I}$ and~$\sigma_{\overline{I}}$.

The second cause for $\chi^2$s that are too large is that the number of samples~$\Nev$ is insufficiently large to guarantee Gaussian statistics for the $I_j$, even after the \vegas~map has fully adapted to the integrand. The threshold for an adequate~$\Nev$ is highly dependent on the integrand. In practice one finds this threshold through trial and error, by looking to see how large~$\Nev$ needs to be in order to obtain stable results and a reasonable~$\chi^2$. Although the statistical error in~$\overline{I}$ can be reduced by increasing either the number of samples~$\Nev$ or the number of iterations~$\Nit$, it is generally better to increase~$\Nev$ while keeping~$\Nit$ just large enough to find the optimal \vegas~map and measure a~$\chi^2$. $\Nit=5$--20 is usually enough.

There is another, related reason for increasing the number of samples~$\Nev$ rather than the number of iterations~$\Nit$. While the estimates~$I_j$ from individual iterations give unbiased estimates of the integral for any~$\Nev$, the weighted sum~$\overline I$ only becomes unbiased when~$\Nev$ is sufficiently large that non-Gaussian effects are negligible. The leading non-Gaussian effect is the correlation between fluctuations in~$I_j$ and~$\sigma_j^2$ (\Eq{correlation}). \red{It introduces a bias in the weighted average that vanishes like $1/\Nev$ with increasing~$\Nev$ and so is usually negligible compared to the statistical uncertainty~$\sigma_{\overline{I}}$, which vanishes more slowly. For example, the bias in~$\overline{I}$ for the second integral discussed above (\Eq{balls}) is only about~$-0.05$\% when~$\Nev=10^5$ (with $\alpha=0.2$). This  is seven times smaller than $\sigma_{j}$, and so is negligible compared to~$\sigma_{\overline{I}}$ unless $\Nit\ge50$. The bias falls to~$-0.03$\% when~$\Nev=2\times10^5$.}

The bias coming from the weighted average can usually be ignored, but it is easy to avoid it completely, if desired. This is done by discarding results from the first~10 or~20 iterations of the \vegas\ algorithm (while the grid is adapting), and then preventing the algorithm from adapting in subsequent iterations (by setting damping parameter~$\alpha=0$). The unweighted average of the latter~$I_j$s provides an unbiased estimate of the integral, and the unweighted average of the~$\sigma_j$s divided by $\sqrt{\Nit}$ gives an estimate of the statistical uncertainty. \red{This technique is useful when the number of samples~$\Nev$ is small, leading to large fluctuations from iteration to iteration in the uncertainties~$\sigma_j$.}

\section{Adaptive Stratified Sampling}
\label{sec:adaptive_stratified_sampling}
The $y$-space (\Eq{y-Dspace}) integrals in the previous section were estimated using Simple Monte Carlo integration. The \vegas\ implementation currently in wide use (``classic \vegas'') improves on this by using {stratified} Monte Carlo sampling in $y$-space rather than Simple Monte Carlo~\cite{vegas}. Given $\Nev$ samples per iteration, the algorithm divides each $y^\mu$~axis ($\mu=1\ldots D$) into 
\begin{equation}
    \Nst = \mathrm{floor}((\Nev / 2)^{1/D}),
    \label{eq:Nst}
\end{equation}
stratifications of width
\begin{equation}
    \Delta y_\mathrm{st} = 1 / \Nst.
\end{equation}
This divides divide $y$-space into $\Nst^D$ hypercubes, each with volume~$\Delta y_\mathrm{st}^D$. The full integral and its variance are the sums of contributions from each hypercube~$h$:
\begin{align}
    I &= \sum_h \Delta I_h \approx \Imc \nonumber \\
    \sigma_I^2 &= \sum_h \sigma_{\Delta I_h}^2 \approx \sigmc^2.
    \label{eq:sumI}
\end{align}
Simple Monte Carlo estimates are made for $\Delta I_h$ (c.f., \Eq{MCIy}) and  $\sigma_{\Delta I_h}^2$ (c.f., \Eq{MCsigIy}) to obtain estimates~$\Imc$ and~$\sigmc^2$ for the total integral and its variance, respectively. The number of integrand samples used is 
\begin{equation}
    n_\mathrm{ev} \equiv \mathrm{floor}(\Nev / \Nst ^ D) \ge 2
\end{equation}
per hypercube.
The standard deviation for a stratified Monte Carlo estimate typically falls with increasing~$\Nst$, potentially as quickly as~$1/\Nst^D\propto 1/\Nev$~\cite{oldref}.

We can improve on the stratification strategy used by classic \vegas\ by allowing the number of integrand samples~$n_h$ used in each hypercube~$h$ to vary from hypercube to hypercube. Then the variance in the Monte Carlo estimate for the integral is 
\begin{equation}
    \sigma_I^2 = \sum_h \frac{\sigma_h^2(Jf)}{n_h}
    \label{eq:strat-sig2}
\end{equation}
where 
\begin{align}
    \sigma_h^2(Jf) &\equiv \Omega_h \int_{\Omega_h}\!d^Dy\,\big(J(y) f(x(y))\big)^2 \nonumber \\
                    &- \Big(\int_{\Omega_h}\!d^Dy\,J(y) f(x(y))\Big)^2
\end{align}
and $\Omega_h$ is the hypercube's volume in $y$-space. Varying the $n_h$ independently, 
constrained by 
\begin{equation}
    \sum_h n_h = \Nev,
\end{equation}
it is easy to show that $\sigma_I^2$ is minimized when 
\begin{equation}
    n_h \propto \sigma_h(Jf).
    \label{eq:optimal-strat}
\end{equation}

The innovation in the new \vegas\ (``\vegasp'') is to redistribute integrand samples across the hypercubes, according to~\Eq{optimal-strat}, after each iteration. The~$\sigma_h(Jf)$  are estimated in an iteration using the integrand samples used to estimate the integral. In this way the distribution of integrand samples across hypercubes is optimized over several iterations, at the same time as the \vegas~map is optimized.

In detail, the algorithm for reallocating samples across hypercubes in \vegasp\ is as follows:
\begin{enumerate}
    \item Choose a somewhat smaller number of stratifications so there are enough samples to allow for significant variation in~$n_h$:
    \begin{equation}
        \Nst = \mathrm{floor}((\Nev/4)^{1/D}).
        \label{eq:Nst_vegasp}
    \end{equation}
    \red{Usually the number stratifications~$\Nst$ is much smaller than the number of increments~$\Ng$ used in the \vegas~map. The algorithm is slightly more stable if 
    $\Ng$~is an integer multiple of~$\Nst$ (or vice versa if $\Nst$~is larger).}

    \item During each iteration accumulate estimates
    \begin{align}
        \sigma_h^2(Jf) &\approx \frac{\Omega_h^2}{n_h} \sum_{y\in\Omega_h} \big(J(y) f(x(y))\big)^2
        \nonumber \\
        &- \Big(\frac{\Omega_h}{n_h} \sum_{y\in\Omega_h} J(y) f(x(y))\Big)^2,
    \end{align}
    for each hypercube using the same samples used to estimate  the integral. 

    \item Introduce a damping parameter~$\beta\ge0$ by replacing $\sigma_h(Jf)$ with
    \begin{equation}
        d_h \equiv \big(\sigma_j(JF)\big)^\beta.
        \label{eq:dh}
    \end{equation}
    Choosing $\beta=1$ corresponds to the optimal distribution (\Eq{optimal-strat}), but a somewhat smaller value can help avoid overreaction to random fluctuations. Setting $\beta=0$ results in $n_h$ values that are all the same\,---\,the stratified sampling becomes non-adaptive, as in classic \vegas. We use $\beta=0.75$ for the examples in this paper.

    \item Recalculate the number of samples for each hypercube,
    \begin{equation}
        n_h = \mathrm{max}\big(2,\Nev {d_h }/{ \sum_{h^\prime}d_{h^\prime}}
        \big),
        \label{eq:nh}
    \end{equation}
    for use in the next iteration. \red{Alternatively, $n_h$ can be set to 
    $\Nev {d_h }/\sum d_{h^\prime} + 2$ which uses more samples but 
    might be more stable. The point of this step is to distribute 
    samples across the hypercubes according to~\Eq{optimal-strat}, while guaranteeing at 
    least 2~samples per hypercube (to allow error estimates).}

\end{enumerate}

The \vegas~map is also updated after each iteration, as described in Section~\ref{sec:iterative-adaptation}. The allocation of integrand samples converges rapidly once the \vegas~map has converged. The optimal \vegas~map for an integrand is independent of the allocation of samples, but reallocating samples according to~\Eq{optimal-strat} can significantly speed the discovery of that optimum because there is better information about the integrand earlier on. The independence of  the optimal \vegas~map from the sample allocation improves the algorithm's stability\,---\,random fluctuations in the sample allocation are unlikely to trigger big changes in the \vegas~map. 

\red{
We mention parenthetically that the original Fortran implementation classic \vegas\ switched to a different form of adaptive stratified sampling than described here when working in low dimensions with lots of samples per iteration. This algorithm replaces the \vegas~map with a similar grid that stratifies $x$~space, but with stratifications concenterated where the uncertainties are largest (rather than where the function is largest); see the 
appendix of Ref.~\cite{vegas} for more details. This algorithm can outperform the classic \vegas\ algorithm in very low dimensions. For example, it is about three times more accurate for the two-dimensional analogue of the integral in Eq.~(\ref{eq:exponentials}) (next section) with 400,000~samples per iteration. The adaptive stratified sampling technique described above, however, is also about three times more accurate than classic \vegas\ for that integral. Typically \vegasp\ does not need this other approach even in low dimensions.
}

\begin{figure}\begin{center}
    \includegraphics[scale=0.9]{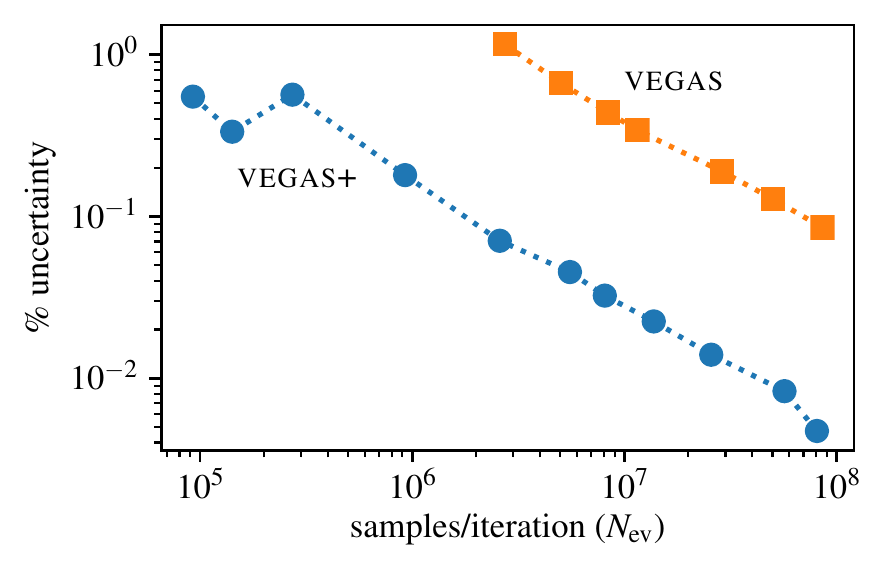}
    \caption{\label{fig:strat} Percent uncertainty ($1\sigma$) in estimates of the integral in Eq.~(\ref{eq:exponentials}) from 30~iterations of classic \vegas\ (top) and \vegasp\ (bottom) is plotted versus the number~$\Nev$ of integrand evaluations (samples) used per iteration. Integral estimates from the first ten iterations are ignored in each case. Damping parameter~$\alpha=0.15$ in both cases. Damping parameter $\beta=0$ for classic \vegas; $\beta=0.75$ for \vegasp. Classic \vegas\ becomes unstable below~$\Nev=3\times10^6$;  \vegasp\ is unstable below $1\times10^5$. The number~$\Nst$ of stratifications per axis used by \vegasp\ varies from~3 to~8  over this range of sample sizes~$\Nev$.}
\end{center}\end{figure}

\subsection{Diagonal Structure}
Adaptive stratified sampling as described in the previous section provides little or no improvement over classic \vegas\ for integrals like those in the previous sections, where the Jacobian from the \vegas~map can flatten the integrand's peaks almost completely and spread them out to fill~$y$-space. It is easy, however, to create integrals  for which the new adaptive stratified sampling makes a big difference. 

Consider, for example,
the eight-dimensional integral
\begin{equation}
    \int\displaylimits_0^1\!d^8x
    \,\sum_{i=1}^3 \er^{-50 \,|\xv - \rv_i|}
    \label{eq:exponentials}
\end{equation}
whose integrand has three narrow peaks along the diagonal of the integration volume, at 
\begin{align}
    \rv_1 &= (0.23, 0.23, 0.23, 0.23, 0.23, 0.23, 0.23, 0.23) \nonumber \\ 
    \rv_2 &= (0.39, 0.39, 0.39, 0.39, 0.39, 0.39, 0.39, 0.39) \nonumber \\ 
    \rv_3 &= (0.74, 0.74, 0.74, 0.74, 0.74, 0.74, 0.74, 0.74).
    \label{eq:random-diag}
\end{align}
The locations  along the diagonal were chosen randomly. Unlike Gaussians, these integrands cannot be factored into a product of separate functions for each direction.

Fig.~\ref{fig:strat} shows that the uncertainties in the integral estimates from classic \vegas\ are 14--19$\times$ larger than those from \vegasp, using the same number of integrand samples~$\Nev$ per iteration. Measured from $\Nev=10^6$, the uncertainty generated by $\Nit$ iterations of the new algorithm falls roughly like
\begin{equation}
    \sigma_I \propto \frac{1}{\sqrt{\Nit}\, \Nev^{0.9}}
    \label{eq:faster}
\end{equation}
with increasing~$\Nit$ and~$\Nev$\,---\,as expected, a larger $\Nev$ is more valuable than a larger~$\Nit$ for a given cost~$\Nit \times \Nev$. \vegasp\ gives reliable results down to $\Nev=10^5$; classic \vegas\ is unusable below $\Nev=3\times10^6$, where it typically misses out one or more of the three peaks.

\begin{figure}\begin{center}
    \includegraphics[scale=0.9]{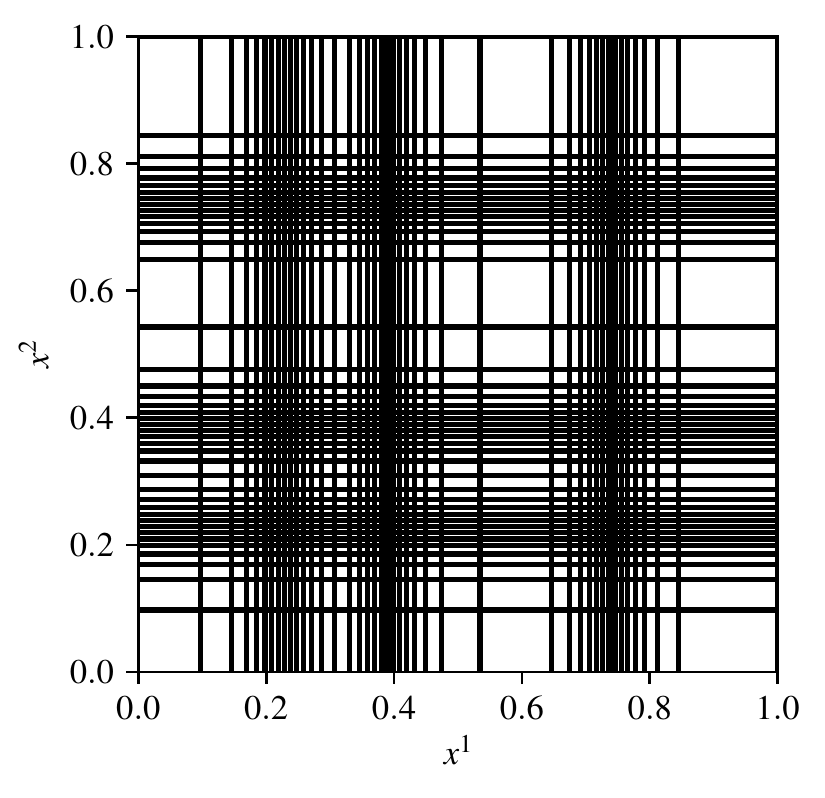}
    \caption{\label{fig:strat2grid} \vegas~map for  the integral in~\Eq{exponentials}. Every 33\textsuperscript{rd} grid line is drawn for axes~$x^1$ and $x^2$; grids for the other axes are the same. }
\end{center}\end{figure}

What makes this integral challenging for classic \vegas\ is the diagonal structure of the integrand. An axis-oriented adaptation strategy, like that used by classic \vegas, will generally have difficulty handling large structures aligned along diagonals. 

\begin{figure}\begin{center}
    \includegraphics[scale=0.9]{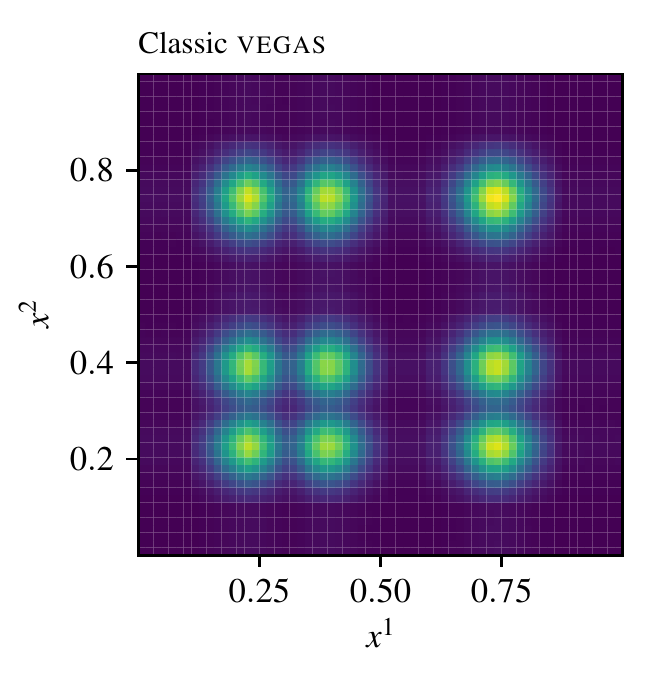}\quad\quad
    \includegraphics[scale=0.9]{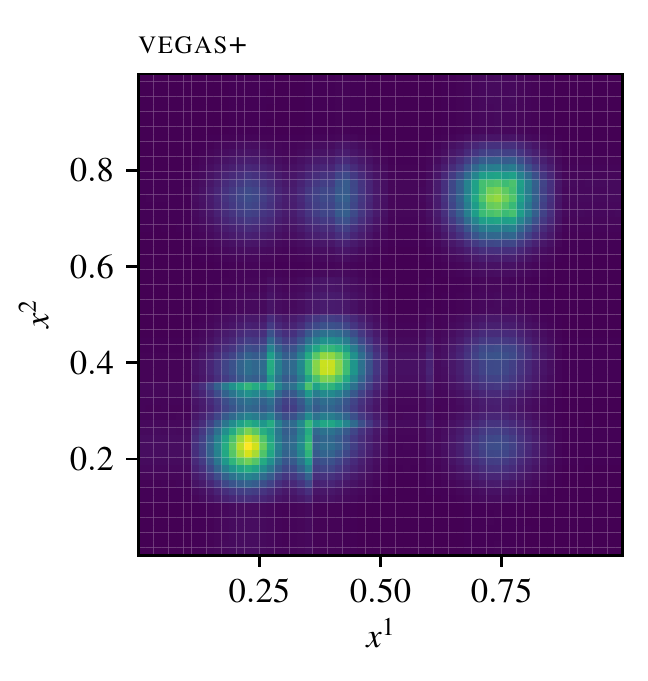}
    \caption{\label{fig:strat2peaks} \red{Histograms showing the distribution of 
    $\Nev\approx10^8$ integrand evaluations across the $x^1$-$x^2$ plane 
    using classic \vegas\ (left) and \vegasp\ (right). The distributions 
    are the same in other directions.}
    }
\end{center}\end{figure}

\begin{figure}\begin{center}
    \includegraphics[scale=0.9]{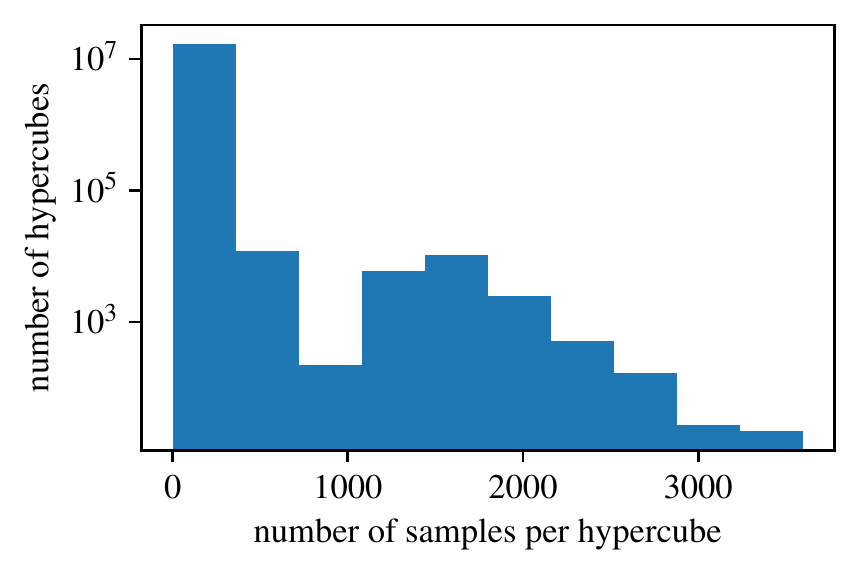}
    \caption{\label{fig:strat2hist} Distribution of integrand samples across hypercubes used by 
    \vegasp\ when evaluating the integral in~\Eq{exponentials}. There are $8^8$~hypercubes and $\Nev\approx10^8$ samples in all.}
\end{center}\end{figure}

The problem for classic \vegas\ is obvious from pictures of the optimal \vegas~map for this integrand (Fig.~\ref{fig:strat2grid}). This grid concentrates integrand  samples at the three peaks on the diagonal, but also at $3^8-3=6558$ additional points, where the integrand is very small:
\begin{align}
    &\rv = (0.39, 0.23, 0.23,\ldots) \nonumber \\ 
    &\rv = (0.74, 0.23, 0.23,\ldots) \nonumber \\ 
    &\rv = (0.23, 0.39, 0.23,\ldots)  \\ 
    &\rv = (0.39, 0.39, 0.23,\ldots) \nonumber \\ 
    &\ldots \nonumber
\end{align}
Integrand samples at these phantom peaks are wasted, greatly reducing the effective number of samples. The adaptive stratification used by \vegasp\ can transfer integration points from the phantom peaks to the real peaks, leading to a much larger effective~$\Nev$. \red{This is evident from the histograms in Fig.~\ref{fig:strat2peaks} which compare the spatial distributions of $\Nev=10^8$ integrand samples using classic \vegas\ (left) and \vegasp\ (right). Classic \vegas\ gives equal attention to peaks and phantoms, while \vegasp\ focuses mostly on the peaks. Fig.~\ref{fig:strat2hist} shows how the samples are distributed across the $8^8$~hypercubes used by \vegasp; more than half of the hypercubes have only 2~samples. Classic \vegas\ uses 2~samples in each of $9^8$~hypercubes.}

\begin{figure}\begin{center}
    \includegraphics[scale=0.9]{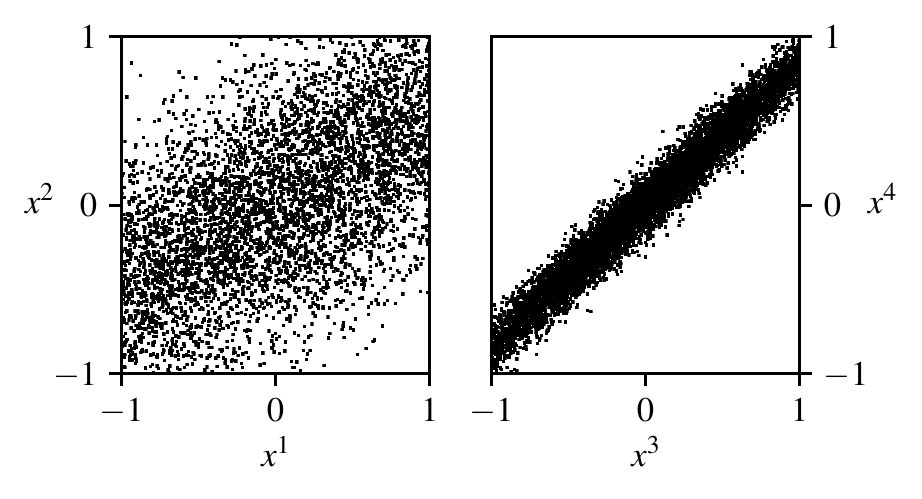}
    \caption{\label{fig:hgaussian} Two views of 10,000 random points distributed with density proportional to the integrand of \Eq{hgaussian}. One view is projected onto the $x^1,x^2$ plane (left) and the other onto the $x^3,x^4$ plane (right).  Correlation coefficients for $x^1,x^2$ and $x^3,x^4$ are~0.866 and~0.986, respectively.}
\end{center}\end{figure}

Another example is the integral 
\begin{equation}
    \int\displaylimits_{-1}^1\!d^4x\,\mathrm{e}^{-\xv^T H^{-1} \xv / 4},
    \label{eq:hgaussian}
\end{equation}
where $H$ is the $4\times4$~Hilbert matrix.\footnote{The $N\times N$ Hilbert matrix has elements $H_{\mu\nu} = 1/(\mu + \nu -1)$ for $\mu,\nu=1\ldots N$. It is famously ill-conditioned.} This integrand has a sharp ridge along an oblique axis (Fig.~\ref{fig:hgaussian}). Classic \vegas\ obtains a 0.25\% accurate result from the last five of seven iterations with $\Nev=4\times10^5$, while \vegasp\ is about 3$\times$~more accurate.

\subsection{Limitations and a Variation}
\label{sec:limitations}
The chief limitation of \vegasp's adaptive stratified sampling is that it requires at least $\Nst=2$ stratifications per direction to have any effect on results. From~\Eq{Nst_vegasp}, this requires  
\begin{equation} 
    \Nev \ge 4 \Nst^D \ge 2^{D+2}
    \label{eq:limitations}
\end{equation} 
integrand evaluations per iteration. While this is not much of a restriction for dimension~$D=5$ or~6, adaptive stratified sampling only turns on when $\Nev\ge1.3\times10^8$ for~$D=25$. This is still manageable but in practice there will be less and less difference between \vegasp\ and classic \vegas\ for higher dimensions. 

\red{
In some situations it is possible to circumvent this restriction, at least partially, by using different numbers~$\Nst^\mu$ of stratifications  in different directions~$\mu$. Most integrands have more structure in some directions than in others. Concentrating stratifications in those directions, with fewer stratifications or none in other directions, allows \vegasp\ to use stratified sampling with smaller values of~$\Nev$. We give an example (with $D=21$) at the end of~\ref{sec:bayesian}.

\begin{figure}\begin{center}
    \includegraphics[scale=0.9]{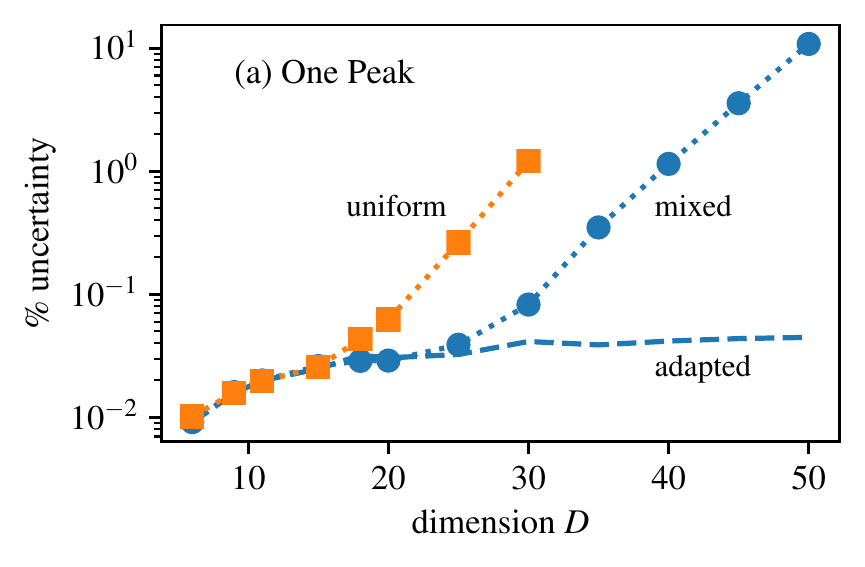}\includegraphics[scale=0.9]{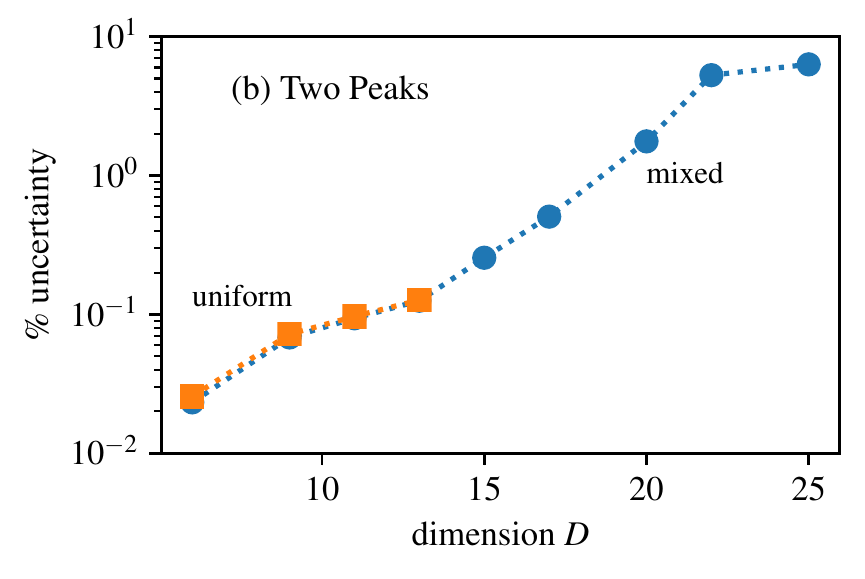}
    \caption{\label{fig:mixed} \red{(a) Percent uncertainty ($1\sigma$) in estimates of the integral in Eq.~(\ref{eq:mixed}) from the last~25 of 50~iterations of \vegasp\ with a uniform stratification (top) and a mixed stratification (bottom) plotted versions the dimension. The damping parameters are $\alpha=0.1$ and~$\beta=0.75$ for the first 25~iterations, while both are set to zero for the remaining iterations (to guarantee unbiased estimates at high dimensions). \vegasp\ is limited to at most 250,000~samples per iteration. The uniform stratification gives unreliable results for~$D>30$. The dashed line shows results from the mixed stratification when the \vegas~map is fully adapted.
    (b) Percent uncertainty for the integral in Eq.~(\ref{eq:twopeaks}) with one fifth as many samples per iteration and five times as many iterations. The uniform stratification gives unreliable results for~$D>13$.}}
\end{center}\end{figure}

Using a mixed set of stratifications rather than a uniform set can be useful in high dimensions even when the integrand does not have structure concentrated in a lower-dimensional subspace. This is illustrated in Fig.~\ref{fig:mixed}(a) where we compare estimates of the integral
\begin{equation}
    \int\displaylimits_0^1\!d^Dx\,\, \er^{-50\,|\xv|}
    \label{eq:mixed}
\end{equation}
for various dimensions up to $D=50$ using two different strategies for stratification. The uniform stratification uses the same number of stratifications in every direction, with the number $\Nst$ determined from Eq.~(\ref{eq:Nst_vegasp}) where $\Nev=2.5\times10^5$. This value for~$\Nst$ is the largest that allows for an average of 4~samples per hypercube given at most $\Nev$~samples per iteration. The mixed stratification uses $\Nst+1$ stratifications for the first~$d$ directions and $\Nst$ stratifications otherwise, where again the value for $d$ is the largest that allows for 4~samples per hypercube. The values for $d$ and $\Nst$ vary with dimension, but $d=15$ and $\Nst=1$ for dimension~$D>15$. So the uniform stratification has only a single hypercube for $D>15$ and adaptive stratified sampling can not be used. The mixed stratification, on the other hand, has $2^{15}$~hypercubes for all~$D>15$. 

The mixed stratification is significantly more accurate for large dimensions $D>15$, and continues working all the way out to $D=50$, while the uniform stratification fails to give useful results above $D=30$. The difference is mostly because the \vegas~map does not have enough iterations to fully adapt in high dimensions. With the mixed stratification, adaptive stratified sampling is still functioning to some considerable extent above $D=15$ and so can help the \vegas~map handle the sharp peak at the origin.

More iterations for adaptation would reduce the uncertainties at large dimension~$D$ for both curves in Fig.~\ref{fig:mixed}(a) (see the dashed line),  since \vegas~maps remain effective for arbitrarily large dimensions once they have converged. More iterations would have little effect, however, on the results in Fig.~\ref{fig:mixed}(b) which use one fifth as many samples per iteration but five times as many iterations (ensuring convergence) to estimate the integral 
\begin{equation}
    \int\displaylimits_0^1\!d^Dx \,\Big( \er^{-4\sum_\mu (x^\mu)^2}
    + \er^{-4\sum_\mu (x^\mu - 1)^2}\Big).
    \label{eq:twopeaks}
\end{equation}
This integrand has peaks at opposite ends of the integration volume's diagonal. The peaks are fairly broad in low dimensions but become increasingly hard for \vegasp\ to discover as the dimension increases. The integrator with the uniform stratification stops working abruptly at $D=14$ where it goes from having $2^{13}$~hypercubes at $D=13$ to only a single hypercube. It is then unable to find both peaks. With the mixed stratification, adaptive stratified sampling can continue beyond $D=13$, with $2^{13}$~hypercubes for all~$D>13$. This partial stratification is enough to stabilize the adaptation of the \vegas~map so that neither peak is lost until much higher dimensions. 

The uncertainties in second example (Fig.~\ref{fig:mixed}(b)) grow quickly with increasing dimension despite the fully adapted \vegas~map. This is because of the exponential growth ($2^{D-2}$) in the number of phantom peaks resulting from the diagonal structure of the integrand (see previous section). Mixed stratification allows adaptive stratified sampling to deal with phantom peaks in 13~directions, which helps but still leaves $2^{D-15}$ phantoms wasting integrand samples. The first example (Fig.~\ref{fig:mixed}(a)) has only a single peak and therefore no phantoms; its uncertainties grow very slowly with~$D$ once the \vegas~map is fully developed (dashed line).

The improvements shown here from mixed stratification are unusually large; for many applications there is little difference between the two strategies. It is probably useful, nevertheless, to use the mixed strategy as the default rather than the uniform strategy described in Section~\ref{sec:adaptive_stratified_sampling} (and used elsewhere in this paper).}

\section{\vegasp\ Hybrids}  
\label{sec:vegasp_hybrids}
In this section we discuss how \vegasp\ can be combined with other algorithms to make hybrid integrators. We look at two strategies. In the first, we use several iterations of \vegasp\ to generate a \vegas~map $x(y)$ that is optimized for the integrand. We then used a different (adaptive) algorithm to evaluate the $y$-space integral~\Eq{y-Dspace}. This strategy, in effect, replaces \vegasp's adaptive stratified sampling with the other algorithm. We illustrate it in Sec.~\ref{sec:miser} by combining \vegasp\ with the widely available \miser\ algorithm, and a variation on that algorithm,~\miserp.

The second strategy employs a Markov Chain Monte Carlo (MCMC) or other peak-finding algorithm to generate a set of samples $\{x, f(x)\}$ of the integrand. These are used to precondition the \vegas~map before integrating. The sample points~$x$ need to cover important regions of the integration volume, but otherwise are unrestricted. The preconditioning makes it easier for \vegasp\ to discover the important regions. In Sec.~\ref{sec:preconditioned} we illustrate this approach with integrands having multiple, narrow peaks. We also compare preconditioned \vegasp\ with a new algorithm optimized for this strategy.

\subsection{\miser\ and \miserp}
\label{sec:miser}
 
\begin{figure}\begin{center}
    \includegraphics[scale=0.9]{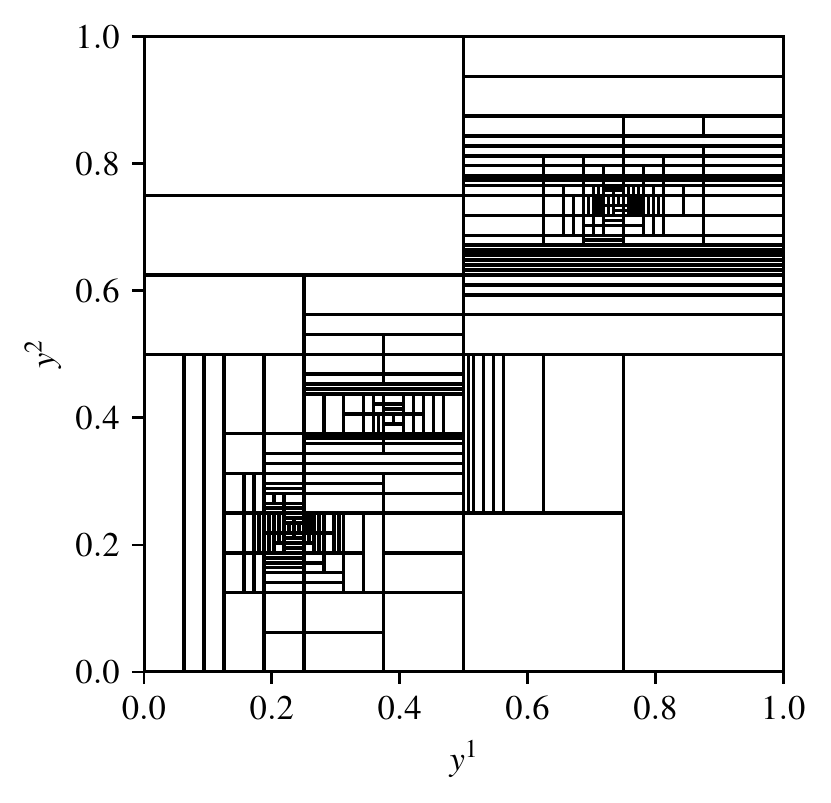}
    \caption{\label{fig:partition} A partition of the integration volume generated by \miser\ for the $D=2$~dimensional version of integral~\Eq{exponentials}. \miser\ used $3\times10^4$ samples to generate this partition and estimate the integral. There are 271~sub-volumes.}
\end{center}\end{figure}

For our first example of a hybrid integrator, we combine \vegasp\ with the \miser\ algorithm~\cite{miser}. \miser\ uses adaptive stratified sampling where the integration volume is recursively partitioned into a large number of rectangular sub-volumes (Fig.~\ref{fig:partition}) to increase the accuracy of the integral's estimate. \miser\ appears to be more effective for integrands with peaks aligned along diagonals than it is for peaks aligned parallel to  integration axes. This is opposite from \vegas~maps, suggesting that the combination might be particularly effective. 

In the following examples, we compare \vegasp\ with \miser\ and with a \vegas-\miser\ hybrid. In each case we run \vegasp\ with various values for the number~$\Nev$ of samples per iteration, and 15~iterations, discarding results from the first~5. We use default values for the damping parameters: $\alpha=0.5$ and $\beta=0.75$. To compare, we run \miser\ with $15\Nev$ samples. The \vegas-\miser\ hybrid uses 5~iterations of \vegasp\ to develop a \vegas~map~$x(y)$ and then estimates the $y$-space integral~\Eq{y-Dspace} using \miser\ with $10\Nev$~integrand samples.

We also compare these algorithms with a variation on \miser, which we call~\miserp. The procedure for \miserp\ is:
\begin{enumerate}
    \item Use \miser\ with half the sample points to generate and save a partition of the integration space optimized for the integrand~$f(x)$. Also save \miser's estimate for 
    \begin{equation}
        \sigma^2_i(f) = \Omega_i \int\displaylimits_{\Omega_i} d^Dx\,f^2(x) 
        - \Big(\int\displaylimits_{\Omega_i} d^Dx\,f(x)\Big)^2
    \end{equation}
    in each sub-volume~$\Omega_i$. Ignore \miser's estimate for the integral.

    \item Distribute the remaining half of the integrand samples across the sub-volumes so that the number~$n_i$ of samples in each sub-volume is proportional to~$\sigma_i(f)$, using the procedure described for~\vegasp\ (Eqs.~(\ref{eq:dh} and~(\ref{eq:nh})).

    \item Estimate the integral in each sub-volume of the partition using Simple Monte Carlo with the number of sample points allocated in the previous step. Add the estimates from each sub-volume to obtain an estimate for the total integral (as in~\Eq{sumI}).
\end{enumerate}
The relation between \miserp\ and \miser\ is similar to that between \vegasp\ and classic \vegas. \miserp\ can also be combined with \vegas~maps, as described above.

\begin{figure}\begin{center}
    \includegraphics[scale=0.875]{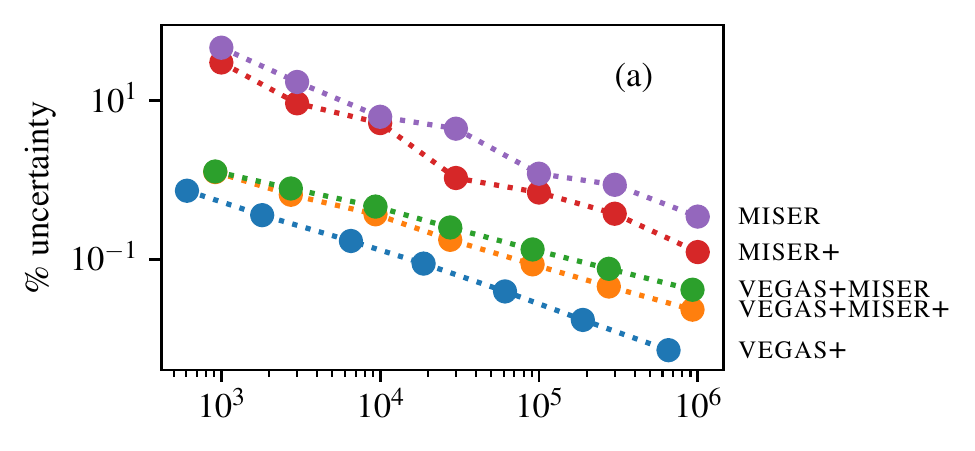}
    
    \includegraphics[scale=0.875]{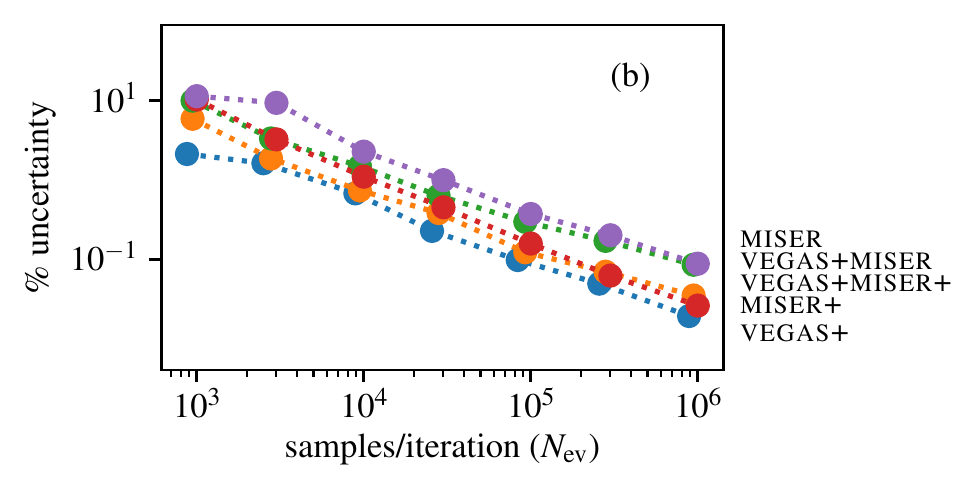}

    \caption{\label{fig:miser} Percent uncertainty ($1\sigma$) in estimates of the integral~\Eq{exponentials4d} with peaks specified by a)~\Eq{parallel}, and b)~\Eq{diagonal}. Results are shown for the last~10 of 15~iterations of \vegasp, with $\Nev$~ integrand evaluations (samples) per iteration. Corresponding results from \miser\ and \miserp\ use $15\Nev$~samples. The \vegas\ hybrids with with \miser\ and \miserp\ use 5~iterations of \vegasp\ to generate a \vegas~map, and then estimate the integral using $10\Nev$ samples with \miser/\miserp.}
\end{center}\end{figure}

We did detailed comparisons for two different integrals. The first is the $D=4$~dimensional integral 
\begin{equation}
    \int\displaylimits_0^1\!d^4x
    \,\sum_{i=1}^3 \er^{-50 \,|\xv - \rv_i|}
    \label{eq:exponentials4d}
\end{equation}
with peaks aligned parallel to the $x^1$~axis:
\begin{align}
    \rv_1 = (0.23, 0.5, 0.5, 0.5) \nonumber \\
    \rv_2 = (0.39, 0.5, 0.5, 0.5) \nonumber \\
    \rv_3 = (0.74, 0.5, 0.5, 0.5) 
    \label{eq:parallel}
\end{align}
The uncertainties in the integral estimates generated by each algorithm for different values of~$\Nev$ are shown in Fig.~\ref{fig:miser}(a). \miser\ is~50--100$\times$ less accurate than~\vegasp. \miser\ is also~1.2--2.7$\times$ less accurate than~\miserp. \vegas~maps are well suited to this integrand, so both \miser\ and \miserp\ see big improvements when used with a \vegas~map. \vegasp\ combined with \miserp\ is 7--38$\times$ more accurate than \miserp\ alone, and only 2--4$\times$ less accurate than~\vegasp\ by itself.

For the second comparison (Fig.~\ref{fig:miser}(b)), we use the same integral but with the peaks aligned along the diagonal of the integration volume:
\begin{align}
    \rv_1 = (0.23, 0.23, 0.23, 0.23) \nonumber \\
    \rv_2 = (0.39, 0.39, 0.39, 0.39) \nonumber \\
    \rv_3 = (0.74, 0.74, 0.74, 0.74) 
    \label{eq:diagonal}
\end{align}
Both \miserp\ and \miser\ are significantly more accurate (4--5$\times$) for this diagonal structure than for the axis-aligned structure of the previous integrand. As discussed above, \vegas~maps are  less effective for diagonal structures, but the combination of \vegasp\ with \miserp\ remains competitive with~\miserp\ alone. \vegasp\ by itself is 4--6$\times$ more accurate than \miser, and 1.5--6$\times$ more accurate than~\miserp. 

We also checked the $D=8$~dimensional versions of these integrals. For the integrand with axis-aligned peaks, \vegasp\ starts working reliably with 500--1000$\times$~fewer integrand samples than \miser\ or \miserp. Using the last 20~of 30~iterations, with $\alpha=0.15$ and $\Nev=10^6$~samples per iteration, \vegasp\ gives 0.03\%-accurate estimates of the integral. \vegas-assisted \miserp\ is 3$\times$ less accurate, while \miserp\ and \miser\ by themselves are both more than 500$\times$ less accurate for a similar number of integrand samples.

The differences are smaller with peaks aligned along the $D=8$~diagonal because both \miser\ and \miserp\ prefer the diagonal structure. \miserp\ is approximately 5$\times$ more accurate than \miser, and only 2--3$\times$ less accurate than~\vegasp\ when~$\Nev$ is between~$5\times10^5$ and~$10^7$. Using a \vegas~map with \miserp\ gives results that vary in precision between \miserp\ and \vegasp, depending upon~$\Nev$. Using \vegasp\ with \miser\ improves on \miser\ but is not as accurate as \miserp.

These experiments suggest that combining \vegasp\ with either \miser\ or \miserp\ is a good idea. Where \vegas~maps are effective, the combination can be much more accurate than~\miser\ or~\miserp\ separately. Where \vegas~maps are less effective, they do not appreciably degrade results compared to the separate algorithms. Typically \miserp\ outperforms \miser\ by factors of~2--5 and so might be the preferred option in combination with~\vegasp. \vegasp\ outperforms all of the other algorithms in all of these tests.

\subsection{Preconditioned Integrators}
\label{sec:preconditioned}
\onlinecite{tq} suggests a different approach to multidimensional integration. They assume that the integrand $f(x)$ has been sampled prior to integration, using MCMC or some other technique. A sample consists of some large number (thousands) of integrand values $\{f(x)\}$  at points $\{x\}$ that are concentrated in regions important to the integral. The samples are used to design an integrator that is customized (preconditioned) for the integrand. The authors describe an algorithm for doing this, but this strategy is also easily implemented using \vegasp.

To illustrate the approach with \vegasp, we consider a dimension~$D=8$ integral whose integrand has three very sharp peaks arrayed along the diagonal of the integration volume:
\begin{equation}
    \int\displaylimits_0^1\!d^8x\,\sum_{i=1}^3 \mathrm{e}^{-10^4 (\xv - \rv_i)^2},
\end{equation}
where the $\rv_i$ are given in \Eq{random-diag}. In this case it is easy to generate random sample points whose density is proportional to the integrand. We use 3000~sample points $\{x\}$. The \vegas~map used by \vegasp\ can be trained on the sample data $\{x,f(x)\}$ before integrating. This is done using the method described in Secs.~\ref{sec:iterative-adaptation} and~\ref{sec:multidimensional}, but with 
\begin{equation}
    d_i^\mu \equiv \frac{1}{n_i^\mu} \sum_{x\in\Delta x_i^\mu} J^2(y(x))\,f^2(x),
\end{equation}
where the sum is over the sample data, $n_i^\mu$ is the number of sample data points falling in increment~$\Delta x_i^\mu$, and $y(x)$ is the inverse of the \vegas~map. The \vegas~map converges quickly as the algorithm is iterated, with the same sample data being reused for each iteration. Here we iterate the algorithm 10~times.

Starting with the preconditioned map, we then run \vegasp\ as usual for $\Nit=8$ iterations, with damping parameter~$\alpha=0$ to prevent further adjustment of the \vegas~map. The 8~iterations allow the stratified sampling algorithm to adapt to whatever structure has not been dealt with by the \vegas~map. As we increase the number~$\Nev$ of integrand evaluations per iteration, we find that \vegasp\ starts to give good results when $\Nev\approx10^4$ to~$10^5$. By $\Nev=10^6$, it is giving 1\%-accurate results. \vegasp\ without the preconditioning is unable to find all three peaks reliably until~$\Nev\approx10^8$ (and classic \vegas\ needs many more evaluations).

We expect a large benefit from preconditioning for extreme multimodal problems like this one. The exact distribution of the sample points~$\{x\}$ is not crucial\,---\, for example, we get more or less the same results above using points drawn from Gaussians that are twice as wide as in the integrand. What matters is that there are enough samples around all of the peaks. In more general problems, we usually do not know \emph{a priori} where the peaks are or how many there are. Peak-finding algorithms or MCMC might be helpful in such cases. As emphasized in \onlinecite{tq}, using preconditioning in this way separates the challenge of finding the integrand's peaks from the challenge of accurate integration once they have been found, allowing us to use different algorithms for the two different tasks, each algorithm optimized for its task.

\begin{figure}\begin{center}
    \includegraphics[scale=0.9]{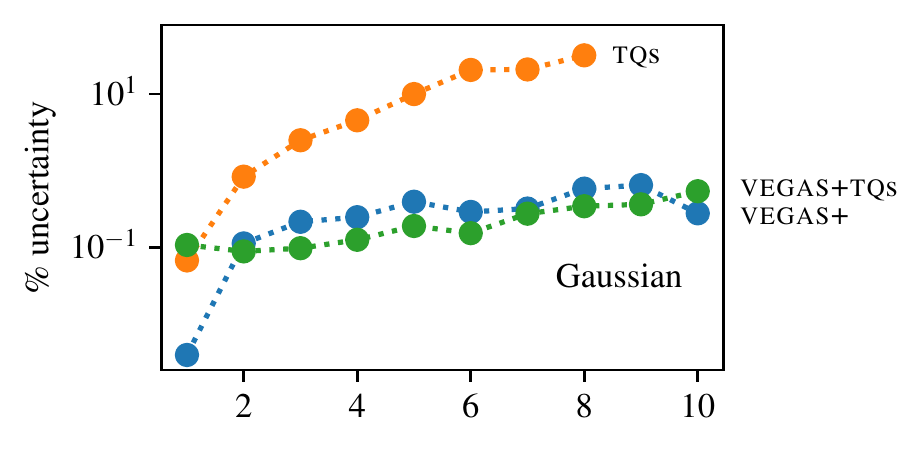}
    \includegraphics[scale=0.9]{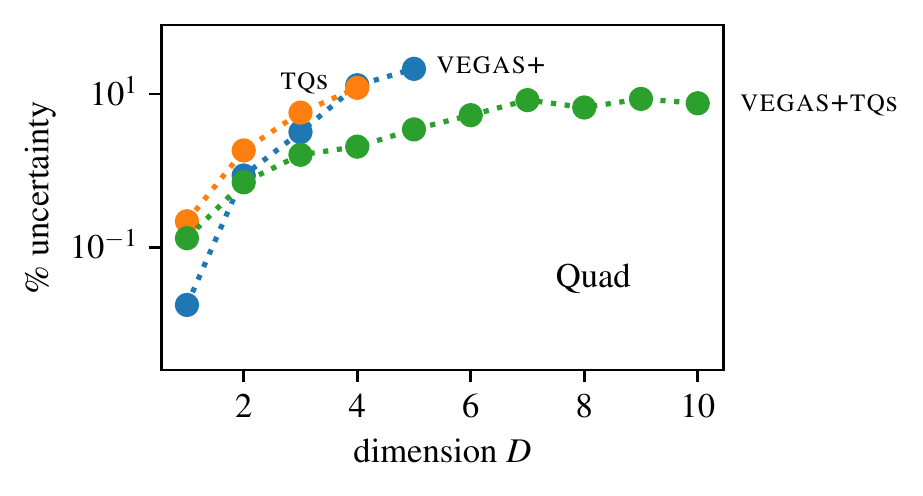}
    \caption{\label{fig:tq} Percent uncertainty ($1\sigma$) in estimates of two integrals from \onlinecite{tq} for dimensions $D=1$--10.  Results are shown for the \vegasp\ (blue) and \tq\ (orange) algorithms, as well as for a hybrid that combines the \vegas~map with~\tq\ (green). All algorithms were limited to 12,000~integrand evaluations in all. The uncertainties are inferred from the interquartile range of the results from 50~repetitions of each integration.}
\end{center}\end{figure}

We compared preconditioned \vegasp\ with one of the algorithms (\tq) from \onlinecite{tq} that is designed for preconditioning. Like \miser, \tq\ recursively subdivides the integration volume into a large number of sub-volumes (c.f., Fig.~\ref{fig:partition}); but \tq\ bases this partitioning on the sample $\{x,f(x)\}$ available before integrating. For our comparison we look at two integrals discussed in \onlinecite{tq}: one has a single narrow Gaussian (same width as in \Eq{easy}) at the center of a $2\times2$~hypercube; the other has four narrow Gaussians (same width as in \Eq{easy}) spread evenly along the diagonal of a $10\times10$~hypercube. The paper labels these problems ``Gaussian'' and ``Quad'', respectively. These are the easiest and hardest integrals considered there. We examined results for dimension~$D=1$--10.

Following \onlinecite{tq}, we limit each algorithm to approximately 12,000~integrand evaluations for these comparisons. The \tq~algorithm uses half~of those samples to divide the integration volume into 2000~sub-volumes. The integral is estimated by doing 3-point Simple Monte Carlo integrals over each sub-volume and summing the results.

\vegasp\ needs far fewer samples to optimize the \vegas~map because optimizing the map for each direction is a separate one-dimensional problem that utilizes all of the data. Here we use 1000~samples to create the \vegas~map. The remaining 11,000~samples are allocated across 4~iterations of \vegasp, again with damping parameter $\alpha=0$. 

Our results are in Fig.~\ref{fig:tq}.
\vegasp\ outperforms \tq\ by more than an order of magnitude on the Gaussian problem in high dimensions, with only modest growth in the errors as the dimension~$D$ increases (the \vegasp\ error is still only 7\% by $D=50$, for example). Not surprisingly, results from the two algorithms are much closer for the Quad problem. Both algorithms become unreliable for dimensions~$D$ greater than three or four\,---\,12,000 integrand samples are too few for higher dimensions.

Fig.~\ref{fig:tq} also shows results for a hybrid approach that combines a \vegas~map with \tq. In the hybrid approach, half of the integrand evaluations are used to create an optimized \vegas~map, as outlined above. Those same samples are then re-used by \tq\ to partition the integration volume to optimize the $y$-space integral~\Eq{y-Dspace} with the optimized \vegas~map~$x(y)$. The $y$-space integral is then calculated summing Simple Monte Carlo estimates of the contributions from each sub-volume in the \tq~partition. The \vegasp\tq\ hybrid gives similar results to \vegasp\ for the Gaussian problem, but outperforms both of the other algorithms on the Quad problem. In particular the hybrid algorithm continues to give usable results even out to dimension~$D=9$--10.

These problems shows how the \vegas~map is effective for dealing with isolated peaks, even when these are arranged along a diagonal of the integration volume. This is because it is able to flatten the peaks. It can't expand the peaks to fill $y$-space when there are multiple peaks along the diagonal, so it is important to combine the map with algorithms, like \vegasp\ and \tq, that can target sub-regions within the $y$-space integration volume.

There are problems, of course, where the \vegas~map is of limited use. The Hilbert-matrix Gaussian in \Eq{hgaussian} is an example. For this problem, neither \tq\ nor the \vegasp\tq\ hybrid can achieve errors smaller than~50\% with only 12,000~samples when $D\ge3$; \vegasp\ gives errors smaller than~1\% for $D=3$.

\section{Conclusions}
\label{sec:conclusions}

In this paper we have  demonstrated how to combine adaptive stratified sampling with classic \vegas's adaptive importance sampling in a new algorithm,~\vegasp. The adaptive stratified sampling makes \vegasp\ far more effective than \vegas\ for dealing with integrands that have multiple peaks or other structure aligned along diagonals of the integration volume. The added computational cost is negligible compared to the cost of evaluating the integrand. \vegasp\ was 2--19$\times$ more accurate than classic \vegas\ in our various examples, with errors that typically fell much faster than $1/\sqrt{\Nev}$ when increasing the number~$\Nev$ of integrand evaluations. 

In Sec.~\ref{sec:vegasp_hybrids}, we showed how to combine \vegasp\ with other algorithms, in effect replacing its adaptive stratified sampling with the other adaptive algorithm. Our experiments with the \miser\ and \tq\ algorithms show that such hybrids can be significantly more accurate than the original algorithms. It would be worthwhile to explore these and other options further.

We also showed (Sec.~\ref{sec:preconditioned}) how to precondition \vegasp\ using integrand samples generated separately from the integrator (e.g., by an MCMC algorithm). Preconditioning can help stabilize \vegasp\ and improve precision, especially for integrands with multiple narrow peaks. In one example, \vegasp\ without preconditioning required more than 100$\times$ as many integrand evaluations as preconditioned \vegasp\ before it began giving reliable results. This again is an area deserving further exploration.

Finally we compared \vegasp\ and MCMC for two Bayes\-ian analyses in \ref{sec:bayesian}, one with $D=3$~parameters and the other with~$D=21$. \vegasp\ was more than 10$\times$ as efficient in both cases. There we discuss why we expect \vegasp\ will often outperform MCMC for small and moderate sized problems.

\section*{Acknowledgments} 
We thank T.\ Kinoshita for sharing his code for the 10\textsuperscript{th}-order QED correction discussed in the first appendix.  This work was supported by the National Science Foundation.

\appendix

\section{Sums and Feynman Diagrams}
\vegasp\ can be used for adaptive multi-dimensional summation, as well as integration. To illustrate,  we show how to use \vegasp\ to correct for the finite space-time volume used in lattice QCD simulations. Such corrections are usually calculated in chiral perturbation theory. A typical example is the contribution from a pion tadpole diagram, which in infinite volume is proportional to 
the integral (in Euclidean space)
\begin{equation}
    I_\pi \equiv \int\displaylimits_{-\infty}^\infty\!d^4k \, f(k),
\end{equation}
where 
\begin{equation}
    f(k) \equiv 1 / \big(k^2 + m^2_\pi\big)^2
\end{equation}
with $m_\pi=0.135$\,GeV. This becomes an infinite, 4-dimensional sum,
\begin{equation}
    S_\pi \equiv (\Delta k)^4 \sum_{k_n} f(k_n),
\end{equation}
when the theory is confined to a box of side~$L$. Here $k^\mu_n=n^\mu \,\Delta k$ with $n^\mu=0, \pm1, \pm2\ldots$
and 
\begin{equation}
    \Delta k = 2\pi / L.
\end{equation}
We take~$L=5$\,fm. The sum is easily converted to an integral:
\begin{equation}
    S_\pi = \int\displaylimits_{-\infty}^\infty\!d^4k \, f(\bar k(k))
\end{equation}
where 
\begin{equation}
    \bar k^\mu(k) \equiv \mathrm{round}(k^\mu / \Delta k)\,\Delta k
\end{equation}
and $\mathrm{round}(x)$ is the nearest integer to~$x$. Then the needed correction can be written,
\begin{align}
    I_\pi - S_\pi &= 16 \int\displaylimits_0^\infty\! d^4k\, \big[f(k) - f(\bar k(k))\big] \\
    &= 16 \int\displaylimits_0^1\! \frac{m_\pi^4\, d^4z}{\prod_\mu (1-z^\mu)^2}\,
    \big[f(k) - f(\bar k(k))\big]
\end{align}
where $k^\mu = m_\pi z^\mu / (1-z^\mu)$.
This integral is ultraviolet finite, unlike the original integrals.

This integral is easy for \vegasp. Using the last~10 of~15 iterations, each with $\Nev=10^3$~samples, gives results with a 7.5\%~uncertainty, which is accurate enough for most practical applications. Here damping parameter~$\alpha=0.5$. Setting $\Nev=10^5$ gives 0.5\%~errors. Classic \vegas\ gives~1.0\% for the same~$\Nev$. 

We also tested \vegasp\ on a 10\textsuperscript{th}-order QED contribution to the muon's magnetic moment. We examined the contributions from the light-by-light diagrams with two vacuum polarization insertions (diagrams~VI(a) in Fig.~6 and Table~IX of \onlinecite{kinoshita2}), using a (Fortran) code for the integrand provided by T.~Kinoshita. The integration is over $D=9$ Feynman parameters. We studied two cases: one where all the fermion-loop particles are muons and the other where they are electrons. The latter has large factors of $\log(m_\mu/m_e)$. We found that \vegasp\ was 3--4$\times$ more accurate than classic \vegas\ in both cases, yielding uncertainties smaller than 0.01--0.02\%, when $\Nev\approx10^8$.

\section{Bayesian Curve Fitting}
\label{sec:bayesian}
\vegasp's ability to find and target narrow high peaks in an integrand makes it well suited for evaluating the integrals used in Bayesian analyses. It can be much faster than other popular methods, such as Markov Chain Monte Carlos (MCMC), when applied to small or medium sized problems. 

\begin{figure}\begin{center}
    \includegraphics[scale=0.9]{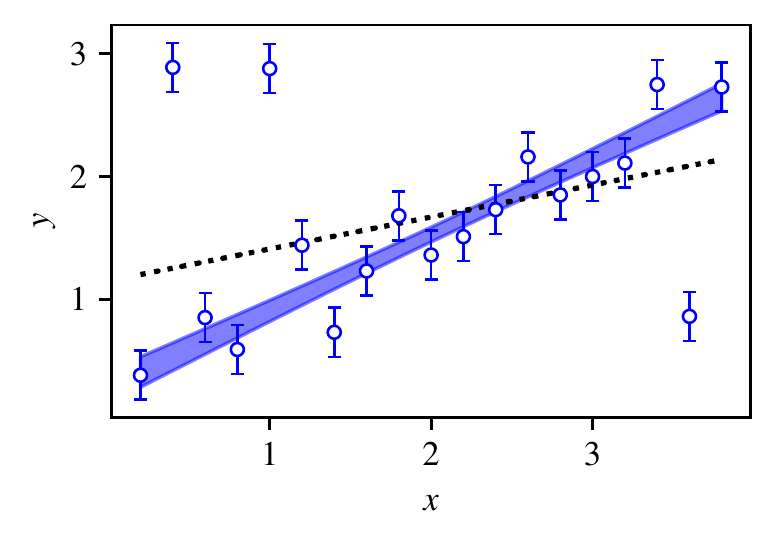}
    \caption{\label{fig:bayes} Bayesian fit (blue band) of data (19~blue data points) with posterior probability \Eq{posterior} and parameters specified by Eqs.~(\ref{eq:bayesvegasp}) and~(\ref{eq:bayescov}). The dotted line shows the fit from a standard least-squares analysis. The band shows the $\pm1\sigma$ range around the best fit line.}
\end{center}\end{figure}

To illustrate a Bayesian analysis, we consider fitting a straight line $p_0 + p_1 x$ to the data in Fig.~\ref{fig:bayes}~\cite{bayes}. The error estimates for several of the points are clearly wrong, so we model the data's probability density as a sum of two Gaussians, one with the nominal width and another with $10\times$~that width:
\begin{align}
    P_\mathrm{data}(y,\sigma_y|\pv,w) &\equiv \frac{(1-w)}{\sqrt{2\pi}\sigma_y}\,\er^{-(y-p_0-p_1 x)^2/2\sigma^2_y}
    \nonumber \\
    &+  \frac{w}{\sqrt{2\pi}\,10\sigma_y}\,\er^{-(y-p_0-p_1 x)^2/200\sigma^2_y}
    \label{eq:posterior}
\end{align}
Here $w$ is the probability of a bad error estimate. Assuming flat priors for the~$p_\mu$ and~$w$, the Bayesian posterior probability density is proportional to
\begin{align}
    f(\pv,w) = P_\mathrm{prior}(\pv,w) \prod_{i=1}^{19} P_\mathrm{data}(y_i,\sigma_{y}|\pv,w),
    \label{eq:prior}
\end{align}
where
\begin{align}
    P_\mathrm{prior}(\pv,w) \propto \Theta(0< w< 1) \prod_{\mu=1}^2 \Theta(-5< p_\mu< 5).
\end{align}

The Bayesian probability distribution is normalized by computing the 3-dimensional integral 
\begin{equation}
    I_0 = \int\displaylimits_{-5}^5\!d^2p \int\displaylimits_0^1\! dw\, f(\pv,w).
\end{equation}
The mean values for the three parameters are calculated from additional integrals,
\begin{align}
    \langle \pv \rangle &= \frac{1}{I_0}  \int\displaylimits_{-5}^5 \!d^2p \int\displaylimits_0^1\! dw\, f(\pv,w)\,\pv\nonumber \\
    \langle w \rangle &= \frac{1}{I_0}  \int\displaylimits_{-5}^5 \!d^2p \int\displaylimits_0^1\! dw\, f(\pv,w)\,w,
    \label{eq:ratios}
\end{align} 
and their covariances from still further integrals:
\begin{align}
    \mathrm{cov_{\pv}} &= \frac{1}{I_0}  \int\displaylimits_{-5}^5 \!d^2p  
    \int\displaylimits_0^1\! dw\, f(\pv,w)\,
    \pv\,\pv^T  - \langle \pv\rangle\,\langle \pv\rangle^T,  \nonumber \\
    \mathrm{var}_w &= \frac{1}{I_0}  \int\displaylimits_{-5}^5 
    \!d^2p  
    \int\displaylimits_0^1\! dw\, f(\pv,w)\,w^2 - \langle w\rangle^2.
    \label{eq:differences}
\end{align}
Additional integrals could provide expectation values~$\langle g(\pv) \rangle$ and/or histograms for arbitrary functions~$g(\pv)$. And so on.

These integrals could be done separately using \vegasp, but it is generally much better to do them simultaneously, using the same sample points~$(\pv,w)$ for all of the integrals. This is because the \vegasp\ errors in the different integrals are then highly correlated, leading to significant cancellations in the errors for ratios like~\Eq{ratios} and differences like~\Eq{differences}. \vegasp\ can estimate the covariances between the estimates of different integrals by summing the covariances coming from each hypercube (estimated using the multivariable generalization of \Eq{sigest}). Given the covariances, it is possible to account for cancellations in the uncertainties associated with ratios, differences, and other combinations of the integration results.

\vegasp\ with 28,000~integrand samples distributed across 15~iterations gives values for the three model parameters that are accurate to 0.06--0.3\% (much more than is needed):
\begin{align}
    \langle\pv\rangle_{\mbox{\vegasp}} &= \big(0.2817(9),\, 0.6224(4)\big) \nonumber \\
    \langle w\rangle_{\mbox{\vegasp}} &= 0.2628(6).
    \label{eq:bayesvegasp}
\end{align}
We drop the first 5~iterations when estimating parameters.
Ignoring correlations between \vegasp\ errors gives results that are 2.5--12$\times$ less accurate.

To compare with MCMC~\cite{emcee}, we generate 150,000~integrand samples with a MCMC and use the last two thirds of those samples to estimate means for the parameters. The results are are 3--4$\times$ less accurate than from \vegasp, despite using more than 5$\times$ as many integrand samples:
\begin{align}
    \langle\pv\rangle_{\mathrm{MCMC},5\times} &= \big(0.2832(29), 0.6215(14)\big) \nonumber \\
    \langle w\rangle_{\mathrm{MCMC},5\times}  &= 0.2638(26)
\end{align}
We estimate the errors for the MCMC simulations by rerunning them several times.

We show the fit line corresponding to the \vegasp\ results for $\langle\pv\rangle$ and 
\begin{equation}
    \mathrm{cov}_{\pv} = 
    \begin{pmatrix}
        0.0179(2) & -0.00675(8) \\
        -0.00675(8) & 0.00318(4)
    \end{pmatrix}
    \label{eq:bayescov}
\end{equation}
in Fig.~\ref{fig:bayes} (blue band). This is more plausible than the fit suggested by a standard least squares analysis (dotted line). Note that the slope and intercept are anti-correlated, with correlation coefficient~$-0.89$. Classic \vegas\ is only somewhat less accurate (40\%) than \vegasp\ for these integrals.

Our MCMC analysis is more than an order of magnitude less efficient than \vegasp\ when computing the means and covariances of the fit parameters. The MCMC's precision is limited by its de-correlation time, the number of MCMC steps required between samples to de-correlate them. This accounts for most of the difference here. The last iteration of \vegasp\ from above, for example, uses $\Nev=1704$~integrand samples to obtain~0.76\% and~0.14\% errors on the intercept and slope, respectively. (The results in~\Eq{bayesvegasp} are from the last 10~iterations and so have  smaller errors.) From \Eq{bayescov}, we can estimate that the same number of uncorrelated samples from the posterior distribution~\Eq{posterior} would give errors of~1.15\% and~0.22\%, respectively. So samples from a perfect Monte Carlo (i.e., de-correlation time equals one step) would be slightly less valuable here than samples from \vegasp, once it has adapted. No general purpose MCMC is perfect, of course. 
The advantage from \vegasp\ samples grows with increasing~$\Nev$, because \vegasp\ errors fall faster than $1/\sqrt{\Nev}$ due to its adaptive stratified sampling (c.f., \Eq{faster}).

We also compared \vegasp\ with MCMC for the more difficult problem where there is a separate value of~$w$ for each data point. Then $w$ becomes a 19-component vector~$\wv$ in the equations above, and the integrals are over 21~variables:~$\pv$ and~$\wv$. \vegasp\ gives results with better than~1\% errors from the last 8~out of 24~iterations, using 162,000~integrand samples in all:
\begin{align}
    \langle\pv\rangle_{\mbox{\vegasp}} &= \big(0.285(2), \, 0.6172(9)\big)    \nonumber \\ 
    \langle\wv\rangle_{\mbox{\vegasp}} &= \big(0.375(3), \, 0.667(3)\,\ldots\big)
\end{align}
An MCMC~analysis with 10$\times$ as many samples gives results that 
are 7--13$\times$ less accurate than from \vegasp:
\begin{align}
    \langle\pv\rangle_{\mathrm{MCMC},10\times} &= \big( 0.291(26), \, 0.6170(77)\big)    \nonumber \\ 
    \langle\wv\rangle_{\mathrm{MCMC},10\times} &= \big( 0.372(25), \, 0.672(19)\,\ldots\big)    
\end{align}
This MCMC analysis used the last third of the samples for estimating parameters.

\red{
Classic \vegas\ and \vegasp\ are the same for this last example since there are only enough integrand samples to allow a single hypercube in 21~dimensions. This follows because the default behavior in \vegasp\ is to use the same number of stratifications in each direction, and $\Nst=2$ is too many (see \Eq{limitations}). We can cut the errors for the slope and intercept in half, however, by using $\Nst^\mu=46$~stratifications in each of the $\pv$~directions, with no stratification ($\Nst^\mu=1$) in the other directions. This can be done with the same number of integrand samples as in the unstratified case.
}

Finally we note that the \vegasp\ results for this problem can be improved (by factors of order~1.5--4) if approximate values for the means of the parameters and their covariance matrix are known ahead of time, for example, from a peak-finding algorithm. Writing the fit parameters as a vector~$\cv$, this is done by first expressing the deviation from the approximate mean~$\cv_0$ in terms of the normalized eigenvectors~$\uv_n$ of the approximate covariance matrix:
\begin{equation}
    \cv \equiv \cv_0 + \sum_n b_n \uv_n.
\end{equation}
The integral is then rewritten as an integral over the coefficients~$b_n$ rather than the components of~$\cv$. This transformation reorients the error ellipse so it is aligned with the integration axes, making it easier for the \vegas~map to adapt around the peak. \onlinecite{supernova} gives an example of this strategy in use.

\end{document}